\documentclass[11pt]{article}
\usepackage{aburas,epsfig}
\usepackage{rotating}

\newcommand{\IM}{{\rm Im}}
\newcommand{\vcb}{|V_{cb}|}
\newcommand{\vtd}{|V_{td}|}
\newcommand{\vub}{|V_{ub}/V_{cb}|}
\newcommand{\vts}{|V_{ts}|}
\newcommand{\vus}{|V_{us}|}

\def\R1{\varepsilon_1}
\def\E8{\varepsilon_8}

\def\eps{\varepsilon}
\def\epe{\varepsilon'/\varepsilon}

\newcommand{\mt}{m_{\rm t}}
\newcommand{\mtb}{\overline{m}_{\rm t}}

\newcommand{\mw}{M_{\rm W}}

\newcommand{\mev}{\, {\rm MeV}}

\newcommand{\bea}{\begin{eqnarray}}
\newcommand{\eea}{\end{eqnarray}}
\newcommand{\bd}{\begin{displaymath}}
\newcommand{\ed}{\end{displaymath}}

\newcommand{\beq}{\begin{equation}}
\newcommand{\eeq}{\end{equation}}
\newcommand{\be}{\begin{equation}}
\newcommand{\ee}{\end{equation}}
\newcommand{\bi}{\begin{itemize}}
\newcommand{\ei}{\end{itemize}}
\newcommand{\ord}{{\cal O}}

\newcommand{\imlt}{\IM\lambda_t}

\begin{document}
\vspace*{4cm}
\title{THE UNITARITY TRIANGLE: 2002 AND BEYOND}

\author{ ANDRZEJ J. BURAS }

\address{Technische Universit{\"a}t M{\"u}nchen, Physik Department,\\ 
 D-85748 Garching, Germany} 

\maketitle\abstracts{We describe the present status of the Unitarity Triangle
and we give
an outlook for its future determinations. We discuss new sets of fundamental 
flavour parameters and comment briefly on new physics beyond the Standard
Model. 
}

\section{CKM Matrix and the Unitarity Triangle}
\setcounter{equation}{0}
The unitary CKM matrix \cite{CAB,KM} connects  the {\it weak
eigenstates} $(d^\prime,s^\prime,b^\prime)$ and 
 the corresponding {\it mass eigenstates} $d,s,b$:
\begin{equation}\label{2.67}
\left(\begin{array}{c}
d^\prime \\ s^\prime \\ b^\prime
\end{array}\right)=
\left(\begin{array}{ccc}
V_{ud}&V_{us}&V_{ub}\\
V_{cd}&V_{cs}&V_{cb}\\
V_{td}&V_{ts}&V_{tb}
\end{array}\right)
\left(\begin{array}{c}
d \\ s \\ b
\end{array}\right)\equiv\hat V_{\rm CKM}\left(\begin{array}{c}
d \\ s \\ b
\end{array}\right).
\end{equation}

Many parametrizations of the CKM
matrix have been proposed in the literature. The classification of different 
parametrizations can be found in \cite{FX1}. While the so called 
standard parametrization \cite{CHAU} should be recommended \cite{PDG} 
for any numerical 
analysis, a generalization of the Wolfenstein parametrization \cite{WO} as 
presented in \cite{BLO} is more suitable for my talk. 
On the one hand it is more transparent than the standard parametrization and 
on the other hand it  satisfies the unitarity 
of the CKM matrix to higher accuracy  than the original parametrization 
in \cite{WO}. Following then the procedure in \cite{BLO} we find 
\be\label{f1}
V_{ud}=1-\frac{1}{2}\lambda^2-\frac{1}{8}\lambda^4, \qquad
V_{cs}= 1-\frac{1}{2}\lambda^2-\frac{1}{8}\lambda^4(1+4 A^2)
\ee
\be
V_{tb}=1-\frac{1}{2} A^2\lambda^4, \qquad
V_{cd}=-\lambda+\frac{1}{2} A^2\lambda^5 [1-2 (\varrho+i \eta)]
\ee
\be\label{VUS}
V_{us}=\lambda+\ord(\lambda^7),\qquad 
V_{ub}=A \lambda^3 (\varrho-i \eta), \qquad 
V_{cb}=A\lambda^2+\ord(\lambda^8)
\ee
\begin{equation}\label{2.83d}
 V_{ts}= -A\lambda^2+\frac{1}{2}A\lambda^4[1-2 (\varrho+i\eta)]
\qquad V_{td}=A\lambda^3(1-\bar\varrho-i\bar\eta)
\end{equation}
where
\begin{equation}\label{2.76}
\lambda, \qquad A, \qquad \varrho, \qquad \eta \, 
\end{equation}
are the Wolfenstein parameters
with $\lambda\approx 0.22$ being an expansion parameter and terms 
$\ord(\lambda^6)$ and higher order terms have been neglected.
A non-vanishing $\eta$ is responsible for CP violation in the SM. It plays 
the role of $\delta_{\rm CKM}$ in the standard parametrization.
Finally, the bared variables in (\ref{2.83d}) are given by
\cite{BLO}
\begin{equation}\label{2.88d}
\bar\varrho=\varrho (1-\frac{\lambda^2}{2}),
\qquad
\bar\eta=\eta (1-\frac{\lambda^2}{2}).
\end{equation}
We emphasize that by definition the expression for
$V_{ub}$ remains unchanged relative to the original Wolfenstein 
parametrization and the
corrections to $V_{us}$ and $V_{cb}$ appear only at $\ord(\lambda^7)$ and
$\ord(\lambda^8)$, respectively.
The advantage of this generalization of the Wolfenstein parametrization
over other generalizations found in the literature 
is the absence of relevant corrections to $V_{us}$, $V_{cd}$, $V_{ub}$ and 
$V_{cb}$ and an  elegant
change in $V_{td}$ which allows a simple generalization of the 
unitarity triangle to higher orders in $\lambda$  
as discussed below. 

Now, the unitarity of the CKM-matrix implies various relations between its
elements. In particular, we have
\begin{equation}\label{2.87h}
V_{ud}^{}V_{ub}^* + V_{cd}^{}V_{cb}^* + V_{td}^{}V_{tb}^* =0.
\end{equation}
Phenomenologically this relation is very interesting as it involves
simultaneously the elements $V_{ub}$, $V_{cb}$ and $V_{td}$ which are
under extensive discussion at present. 
The relation (\ref{2.87h})  can be
represented as a ``unitarity'' triangle in the complex 
$(\bar\varrho,\bar\eta)$ plane. 
One can construct additional five unitarity triangles \cite{Kayser} 
corresponding to other unitarity relations,
but I do not have space to discuss them here.

Noting that to an excellent accuracy $V_{cd}^{}V_{cb}^*$ is real with
$| V_{cd}^{}V_{cb}^*|=A\lambda^3+\ord(\lambda^7)$ and
rescaling all terms in (\ref{2.87h}) by $A \lambda^3$ 
we indeed find that the relation (\ref{2.87h}) can be represented 
as the triangle 
in the complex $(\bar\varrho,\bar\eta)$ plane 
as shown in fig.~\ref{fig:utriangle}. Let us collect useful formulae related 
to this triangle:

\begin{figure}[hbt]
\vspace{0.10in}
\centerline{
\epsfysize=2.1in
\epsffile{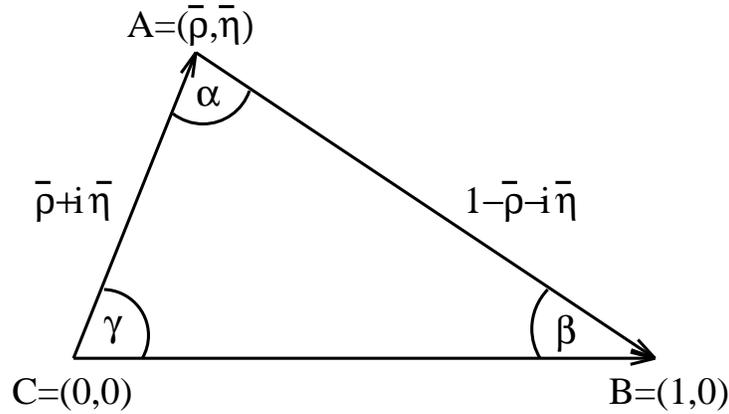}
}
\vspace{0.08in}
\caption{Unitarity Triangle.}\label{fig:utriangle}
\end{figure}

\bi
\item
We can express $\sin(2\phi_i$), $\phi_i=
\alpha, \beta, \gamma$, in terms of $(\bar\varrho,\bar\eta)$. In particular:
\begin{equation}\label{2.90}
\sin(2\beta)=\frac{2\bar\eta(1-\bar\varrho)}{(1-\bar\varrho)^2 + \bar\eta^2}.
\end{equation}
\item
The lengths $CA$ and $BA$ to be denoted by $R_b$ and $R_t$,
respectively, are given by
\begin{equation}\label{2.94}
R_b \equiv \frac{| V_{ud}^{}V^*_{ub}|}{| V_{cd}^{}V^*_{cb}|}
= \sqrt{\bar\varrho^2 +\bar\eta^2}
= (1-\frac{\lambda^2}{2})\frac{1}{\lambda}
\left| \frac{V_{ub}}{V_{cb}} \right|,
\end{equation}
\begin{equation}\label{2.95}
R_t \equiv \frac{| V_{td}^{}V^*_{tb}|}{| V_{cd}^{}V^*_{cb}|} =
 \sqrt{(1-\bar\varrho)^2 +\bar\eta^2}
=\frac{1}{\lambda} \left| \frac{V_{td}}{V_{cb}} \right|.
\end{equation}
\item
The angles $\beta$ and $\gamma=\delta_{\rm CKM}$ of the unitarity triangle 
are related
directly to the complex phases of the CKM-elements $V_{td}$ and
$V_{ub}$, respectively, through
\beq\label{e417}
V_{td}=|V_{td}|e^{-i\beta},\quad V_{ub}=|V_{ub}|e^{-i\gamma}.
\eeq
\item
The unitarity relation (\ref{2.87h}) can be rewritten as
\be\label{RbRt}
R_b e^{i\gamma} +R_t e^{-i\beta}=1~.
\ee
\item
The angle $\alpha$ can be obtained through the relation
\beq\label{e419}
\alpha+\beta+\gamma=180^\circ~.
\eeq
\ei

Formula (\ref{RbRt}) shows transparently that the knowledge of
$(R_t,\beta)$ allows to determine $(R_b,\gamma)$ through 
\be\label{VUBG}
R_b=\sqrt{1+R_t^2-2 R_t\cos\beta},\qquad
\cot\gamma=\frac{1-R_t\cos\beta}{R_t\sin\beta}.
\ee
Similarly, $(R_t,\beta)$ can be expressed through $(R_b,\gamma)$:
\be\label{VTDG}
R_t=\sqrt{1+R_b^2-2 R_b\cos\gamma},\qquad
\cot\beta=\frac{1-R_b\cos\gamma}{R_b\sin\gamma}.
\ee
These relations are remarkable. They imply that the knowledge 
of the coupling $V_{td}$ between $t$ and $d$ quarks allows to deduce the 
strength of the corresponding coupling $V_{ub}$ between $u$ and $b$ quarks 
and vice versa.

The triangle depicted in fig. \ref{fig:utriangle}, $|V_{us}|$ 
and $\vcb$ give the full description of the CKM matrix. 
Looking at the expressions for $R_b$ and $R_t$, we observe that within
the SM the measurements of four CP
{\it conserving } decays sensitive to $|V_{us}|$, $|V_{ub}|$,   
$|V_{cb}|$ and $|V_{td}|$ can tell us whether CP violation
($\bar\eta \not= 0$) is predicted in the SM. 
This fact is often used to determine
the angles of the unitarity triangle without the study of CP-violating
quantities.

\section{The Special Role of \boldmath{$|V_{us}|$}, \boldmath{$|V_{ub}|$}
and \boldmath{$|V_{cb}|$}}
What do we know about the CKM matrix and the unitarity triangle on the
basis of {\it tree level} decays? 
Here the semi-leptonic K and B decays play the decisive role. 
The present situation can be summarized roughly by 
\begin{equation}\label{vcb}
|V_{us}| = \lambda =  0.221 \pm 0.002\,
\quad\quad
\vcb=(40.6\pm0.8)\cdot 10^{-3},
\end{equation}
\begin{equation}\label{v13}
\frac{|V_{ub}|}{\vcb}=0.089\pm0.008, \quad\quad
|V_{ub}|=(3.63\pm0.32)\cdot 10^{-3}.
\end{equation}
implying
\be
 A=0.83\pm0.02,\qquad R_b=0.39\pm 0.04~.
\ee
The errors given here look a bit aggressive and should not be considered 
as giving ranges for the quantities in question. They indicate rather 
standard deviations. See \cite{BUPAST} for more details.
There is an impressive work done by theorists and experimentalists hidden
behind these numbers that
are in the ball park of various analyses present in the 
literature. A very incomplete list of references is given in 
\cite{VCB,ref:ckmworkshop}. 
See also the relevant articles in \cite{PDG}.
Detailed discussions of these analyses with 
possibly updated 
values should be available soon \cite{CERNCKM}.
In particular the very recent analysis of $\vus$ \cite{VUS} gives 
$\vus=0.2241\pm 0.0036.$

\begin{figure}[hbt]
\vspace{-0.10in}
\centerline{
\epsfysize=2.0in
\epsffile{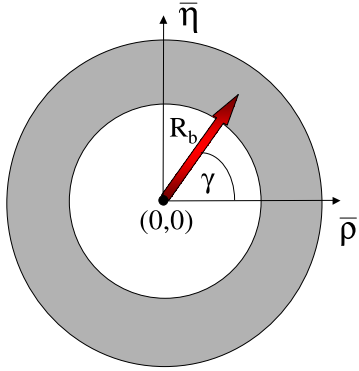}
}
\vspace{0.02in}
\caption[]{``Unitarity Clock".
\label{L:2}}
\end{figure}

The information given above tells us only that the apex $A$ of the unitarity 
triangle lies in the band shown in fig.~\ref{L:2}. 
While this information appears at first sight to be rather limited, 
it is very important for the following reason. As $|V_{us}|$, $\vcb$, 
 $|V_{ub}|$ and consequently $R_b$ are determined here from tree level 
decays, their
values given above are to an excellent accuracy independent of any 
new physics contributions. They are universal fundamental 
constants valid in any extention of the SM. Therefore their precise 
determinations are of utmost importance. 
In order to answer the question where
the apex $A$ lies on the ``unitarity clock'' in fig.~\ref{L:2} we have to 
look at other decays. Most promising in this respect are the so-called 
``loop induced'' decays and transitions and CP-violating B decays. 
These decays are sensitive to the angles $\beta$ and $\gamma$ as well as 
to the length $R_t$ and measuring only one of these three quantities allows to 
find the unitarity triangle provided the universal $R_b$ is known.

Of course any pair among $(R_t,\beta,\gamma)$ is sufficient to construct 
the UT without any knowledge of $R_b$. Yet the special role of $R_b$ among
these variables lies in its universality whereas the other three variables 
are generally sensitive functions of possible new physics contributions.
This means that assuming three generation unitarity 
of the CKM matrix and that the SM is a part of a bigger 
theory, the apex of the unitarity triangle has to be eventually placed
on the unitarity clock with the radius $R_b$ obtained from tree level decays.
That is even 
if using SM expressions for loop induced processes, $(\bar\varrho,\bar\eta)$
would be found outside the unitarity clock, the corresponding expressions 
of the grander theory must include appropriate new contributions so that 
the apex of the unitarity triangle is shifted back to the band in  
fig.~\ref{L:2}. In the case of CP asymmetries this could be achieved by 
realizing that the measured angles $\alpha$, $\beta$ and $\gamma$ are not
the true angles of the unitarity triangle but sums of the true angles and 
new complex phases present in extentions of the SM. The better $R_b$ is known,
the thiner the band in fig.~\ref{L:2} will be, 
selecting in this manner efficiently the correct theory. On the other hand 
as the 
the branching ratios for rare and CP-violating decays depend sensitively
on the parameter $A$, the precise knowledge of $\vcb$ is also very important.

\section{Standard Analysis of the Unitarity Triangle}\label{UT-Det}
After these general remarks let us concentrate on
the standard analysis of the Unitarity Triangle within the SM. 
The so-called standard analysis of the UT of fig. \ref{fig:utriangle} 
involves the values of $\vus$, 
$\vcb$ and $\vub$ extracted from tree level decays, the parameter 
$\varepsilon_K$ that describes the indirect CP violation in $K_L\to\pi\pi$ 
decays and 
the differences of mass eigenstates $\Delta M_{d,s}$ in the 
$B^0_d-\bar B^0_d$ and $B^0_s-\bar B^0_s$ systems. 
Setting $\lambda=\vus$, the analysis
proceeds in the following five steps:

{\bf Step 1:}

{}From  $b\to c$ transition in inclusive and exclusive 
leading B-meson decays
one finds $\vcb$ and consequently the scale of the unitarity triangle:
\begin{equation}
\vcb\quad \Longrightarrow\quad\lambda \vcb= \lambda^3 A~.
\end{equation}

{\bf Step 2:}

{}From  $b\to u$ transition in inclusive and exclusive $B$ meson decays
one finds $\vub$ and consequently using (\ref{2.94}) 
the side $CA=R_b$ of the unitarity triangle:
\begin{equation}\label{rb}
\left| \frac{V_{ub}}{V_{cb}} \right|
 \quad\Longrightarrow \quad R_b=\sqrt{\bar\varrho^2+\bar\eta^2}=
4.41 \cdot \left| \frac{V_{ub}}{V_{cb}} \right|~.
\end{equation}

{\bf Step 3:}

\begin{figure}[hbt]
  \vspace{0.10in} \centerline{
\begin{turn}{-90}
  \mbox{\epsfig{file=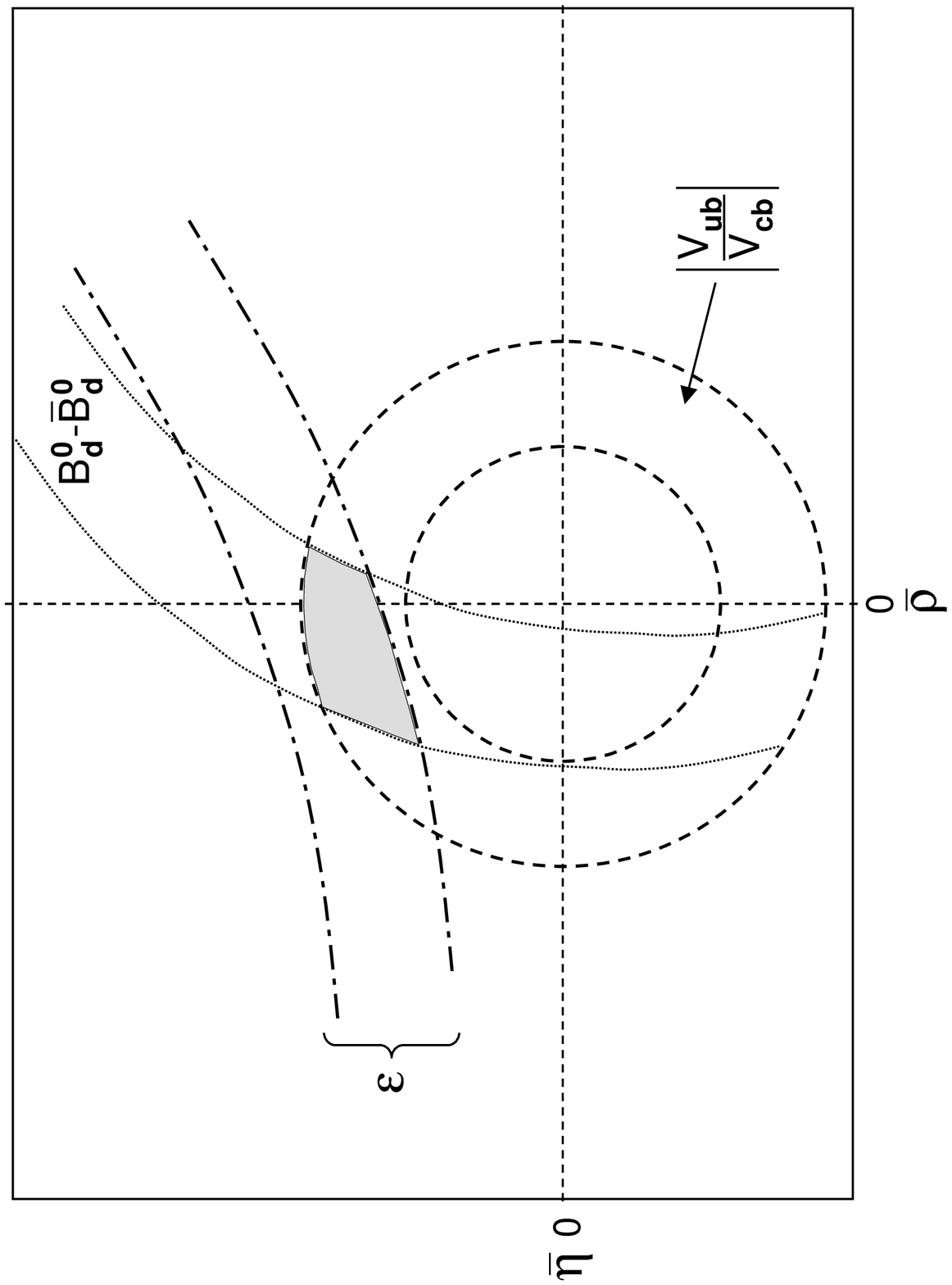,width=0.5\linewidth}}
\end{turn}
} \vspace{-0.18in}
\caption[]{Schematic determination of the Unitarity Triangle.
\label{L:10}}
 \end{figure}

{}From the experimental value of $\varepsilon_K$  
and the standard calculation of box diagrams describing $K^0-\bar K^0$ mixing
one derives including QCD corrections \cite{Erice} 
the constraint $(\lambda = 0.221)$ \cite{WARN}
\begin{equation}\label{100}
\bar\eta \left[(1-\bar\varrho) A^2 \eta_2 S_0(x_t)
+ P_c(\varepsilon) \right] A^2 \hat B_K = 0.214,
\end{equation}
where
\begin{equation}\label{102}
P_c(\varepsilon) = 
\left[ \eta_3 S_0(x_c,x_t) - \eta_1 x_c \right] \frac{1}{\lambda^4},
\qquad
x_i=\frac{m^2_i}{\mw^2}.
\end{equation}
 $S_0(x_t)$ and $S_0(x_t,x_c)$ are known functions \cite{IL,BSS} and 
$P_c(\varepsilon)=0.28\pm0.05$ \cite{Nierste} summarizes the contributions
of box diagrams with two charm quark exchanges and the mixed 
charm-top exchanges. $\hat B_K$ is a non-perturbative parameter that
represents the relevant hadronic matrix element, the main uncertainty 
in (\ref{100}).
The short-distance QCD effects are described through the correction
factors $\eta_1$, $\eta_2$, $\eta_3$.
The NLO values of $\eta_i$ with an updated $\eta_1$ \cite{Nierste} 
are given as follows \cite{HNa,BJW90,HNb}:
\begin{equation}
\eta_1=1.45\pm 0.38,\qquad
\eta_2=0.57\pm 0.01,\qquad
  \eta_3=0.47\pm0.04~.
\end{equation}
As illustrated in fig. \ref{L:10}, equation (\ref{100}) specifies 
a hyperbola in the $(\bar \varrho, \bar\eta)$
plane.
The position of the hyperbola depends on $\mt$, $|V_{cb}|=A \lambda^2$
and $\hat B_K$. With decreasing $\mt$, $|V_{cb}|$ and $\hat B_K$ the
$\eps_K$-hyperbola moves away from the origin of the
$(\bar\varrho,\bar\eta)$ plane. 

{\bf Step 4:}

{}From the observed $B^0_d-\bar B^0_d$ mixing parametrized by $\Delta M_d$ 
and the standard calculation of box diagrams describing this mixing,
the side $AB=R_t$ of the unitarity triangle can be determined:
\begin{equation}\label{106}
 R_t= \frac{1}{\lambda}\frac{|V_{td}|}{\vcb} = 0.86 \cdot
\left[\frac{|V_{td}|}{7.8\cdot 10^{-3}} \right] 
\left[ \frac{0.041}{\vcb} \right]
\end{equation}
with
\begin{equation}\label{VT}
\vtd=
7.8\cdot 10^{-3}\left[ 
\frac{230\mev}{\sqrt{\hat B_{B_d}}F_{B_d}}\right]
\left[\frac{167~GeV}{\mtb(\mt)} \right]^{0.76} 
\left[\frac{\Delta M_d}{0.50/{\rm ps}} \right ]^{0.5} 
\sqrt{\frac{0.55}{\eta_B}}.
\end{equation}
Here $\eta_B=0.55\pm0.01$ summarizes the NLO QCD corrections 
\cite{BJW90,UKJS} and $F_{B_d}\sqrt{\hat B_{B_d}}$ describes the 
relevant hadronic matrix element. $\mtb(\mt)=(167\pm 5)$ GeV.
Note that $R_t$ suffers from additional uncertainty in $\vcb$,
which is absent in the determination of $\vtd$ this way. 
The constraint in the $(\bar\varrho,\bar\eta)$ plane coming from
this step is illustrated in fig.~\ref{L:10}.

{\bf Step 5:}

{}The measurement of $B^0_s-\bar B^0_s$ mixing parametrized by $\Delta M_s$
together with $\Delta M_d$  allows to determine $R_t$ in a different
manner:
\be\label{Rt}
R_t=0.88~\left[\frac{\xi}{1.18}\right] \sqrt{\frac{18.0/ps}{\Delta M_s}} 
\sqrt{\frac{\Delta M_d}{0.50/ps}},
\qquad
\xi = 
\frac{\sqrt{\hat B_{B_s}}F_{B_s} }{ \sqrt{\hat B_{B_d}}F_{B_d}}.
\ee
One should 
note that $\mt$ and $|V_{cb}|$ dependences have been eliminated this way
 and that $\xi$ should in principle 
contain much smaller theoretical
uncertainties than the hadronic matrix elements in $\Delta M_d$ and 
$\Delta M_s$ separately.  

The main uncertainties in this analysis originate in the theoretical 
uncertainties in the non-perturbative parameters $\hat B_K$ and 
$\sqrt{\hat B_d}F_{B_d}$ and to a lesser extent in $\xi$ \cite{LATT}: 
\be
\hat B_K=0.86\pm0.15, \qquad  \sqrt{\hat B_d}F_{B_d}=(235^{+33}_{-41})~MeV,
\qquad \xi=1.18^{+0.13}_{-0.04}~.
\ee
The significant uncertainty in $\xi$ is disturbing \cite{KRRY} and should be 
clarified.
Also the uncertainty due to $\vub$ in step 2 should certainly be decreased.
The QCD sum rules results for 
the parameters in question are similar and can be found in \cite{Jamin}. 
Finally \cite{DMDDMS}
\be
\Delta M_d=(0.503\pm0.006)/{\rm ps}, \qquad 
\Delta M_s>14.4/{\rm ps}~~ {\rm at }~~ 95\%~{\rm C.L.}
\ee 
\section{The Angle \boldmath{$\beta$} from \boldmath{$B\to \psi K_S$}}
One of the highlights of the year 2002 were the considerably improved 
measurements of 
$\sin2\beta$ by means of the time-dependent CP asymmetry
in $B^0_d(\bar B^0_d)\to \psi K_S$ decays 
\be\label{asy}
a_{\psi K_S}(t)\equiv -a_{\psi K_S}\sin(\Delta M_d t)=
-\sin 2 \beta \sin(\Delta M_d t)
\ee
The most recent measurements of $a_{\psi K_S}$ from
the BaBar \cite{BaBar} and Belle \cite{Belle} Collaborations imply
\begin{displaymath}\label{sinb}
(\sin 2\beta)_{\psi K_S}=\left\{
\begin{array}{ll}
0.741\pm 0.067 \, \mbox{(stat)} \pm0.033 \, \mbox{(syst)} & \mbox{(BaBar)}\\
0.719\pm 0.074 \, \mbox{(stat)} \pm0.035  \, \mbox{(syst)} & \mbox{(Belle).}
\end{array}
\right.
\end{displaymath}
Combining these results with earlier measurements by CDF 
$(0.79^{+0.41}_{-0.44})$, ALEPH $(0.84^{+0.82}_{-1.04}\pm 0.16)$ and OPAL 
gives the grand average \cite{NIR02}
\be
(\sin 2\beta)_{\psi K_S}=0.734\pm 0.054~.
\label{ga}
\ee
This is a mile stone in the field of CP violation and in the tests of the
SM as we will see in a moment. Not only violation of this symmetry has been 
confidently established 
in the B system, but also its size has been measured very accurately.
Moreover in contrast to the constraints of section 3, the determination of 
the angle $\beta$ in this manner does not practically suffer from any hadronic 
uncertainties.
\section{Unitarity Triangle 2002}
\begin{figure}[htb!]
\begin{center}
{\epsfig{figure=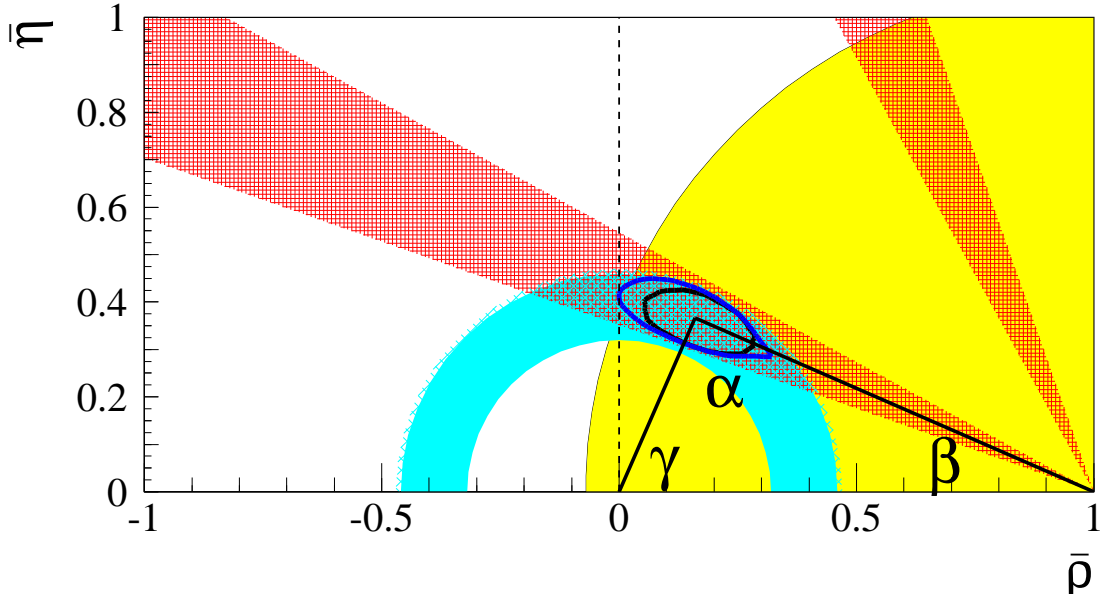,height=9cm}}
\caption[]{The allowed 95$\%$ regions in the 
$(\bar\varrho,\bar\eta)$ plane in the SM (narrower region) and in the 
MFV models (broader region) from \cite{BUPAST}.
The individual 95$\%$ regions for the constraint from 
$\sin 2 \beta$, $\Delta M_s$ and $R_b$ are also shown.  
The results are
obtained using the fit procedure described in \cite{C00}}
\label{fig:figmfv}
\end{center}
\end{figure}

We are now in the position to combine all these constraints in order to 
construct the unitarity triangle and determine various quantities of interest.
In this context the important issue is the error analysis of these formulae, 
in particular the treatment of theoretical uncertainties. In the 
literature five different methods are commonly used: 
Gaussian approach \cite{Gaus}, 
Bayesian approach \cite{C00}, frequentist approach \cite{FREQ}, 
$95\%$ C.L. scan method \cite{SCAN95} and the simple 
(naive) scanning within one standard deviation as used 
by myself in the past and a few distinguished colleagues of my generation. 
For the PDG analysis see \cite{PDG} and Kleinknechts talk.
A critical comparision of these methods will appear soon \cite{CERNCKM}.
Recently I have been converted to the Bayesian approach.
Consequently,
in fig.~\ref{fig:figmfv} the result of an analysis in collaboration 
with Parodi and Stocchi \cite{BUPAST} that uses this approch is shown. 
The allowed region for $(\bar\varrho,\bar\eta)$ 
is the area inside the smaller ellipse.
We observe that the region
$\bar\varrho<0$ is disfavoured by the lower bound on
$\Delta M_s$.
It is clear
from this figure that $\Delta M_s$ is a very important
ingredient in this analysis and that the measurement of $\Delta M_s$
giving $R_t$ through (\ref{Rt}) will have a large impact
on the plot in fig.~\ref{fig:figmfv}. 
Other relatively recent analyses of the UT in the SM can be found in 
\cite{Gaus,C00,FREQ,bologna}.

The ranges for various quantities that result from this analysis 
are given in the last column of table~\ref{mfv}. The first column 
will be discussed at the end of my talk. The results in this table 
follow from five steps of section 3 and the direct measurement of 
$\sin 2\beta$ in (\ref{ga}). They imply in particular an impressive 
precision on the angle $\beta$:
\begin{equation}\label{TOT}
(\sin 2\beta)_{\rm tot}=0.725\pm 0.033, \qquad 
 \beta=(23.2 \pm 1.4)^\circ~.
\end{equation}
On the other hand $(\sin 2\beta)_{\rm ind}$ obtained by using only 
the five steps of section 3 is found to be 
\cite{BUPAST}
\be
(\sin 2\beta)_{\rm ind}=0.715^{+0.055}_{-0.045}
\label{ind}
\ee
demonstrating an excellent agreement (see also fig.~\ref{fig:figmfv}) 
between the direct measurement in (\ref{ga})
and the standard analysis of
the unitarity triangle within the SM.
This gives a strong indication that the CKM matrix is very likely 
the dominant source of CP violation in flavour violating decays.
In order to be sure whether this is indeed the case other theoretically
clean quantities have to be measured. In particular the angle $\gamma$ 
that is more sensitive to new physics contributions than $\beta$.
In this context the measurement of the ratio  $\Delta M_s/\Delta M_d$
will play an important role as for a fixed value of $\sin 2\beta$, 
the extracted value for $\gamma$ is a sensitive function of 
$\Delta M_s/\Delta M_d$.

\begin{table}[htb!]
\begin{center}
\begin{tabular}{|c|c|c|}
\hline
  Strategy       &               UUT              &            SM                \\
  $\bar {\eta}$  &         0.369 $\pm$ 0.032      &     0.357 $\pm$ 0.027        \\ 
                 &   (0.298-0.430) [0.260-0.449]  & (0.305-0.411)  [0.288-0.427] \\
  $\bar {\rho}$  &         0.151 $\pm$ 0.057      &     0.173 $\pm$ 0.046        \\
                 &   (0.034-0.277) [-0.023-0.358]  & (0.076-0.260)  [0.043-0.291] \\
  $\sin 2\beta$  &   0.725 $^{+0.038}_{-0.028}$   &  0.725 $^{+0.035}_{-0.031}$  \\
                 &   (0.661-0.792) [0.637-0.809]  & (0.660-0.789)  [0.637-0.807] \\
  $\sin 2\alpha$ &        0.05 $\pm$ 0.31         &   -0.09 $\pm$ 0.25           \\
                 &   (-0.62-0.60)  [-0.89-0.78]   & (-0.54-0.40)   [-0.67-0.54]  \\
  $\gamma$       &         67.5 $\pm$ 9.0         &    63.5 $\pm$ 7.0            \\
  (degrees)      &   (48.2-85.3)   [36.5-93.3]    & (51.0-79.0)    [46.4-83.8]   \\
  $R_b$          &         0.404 $\pm$ 0.023      &   0.400 $\pm$ 0.022          \\
                 &   (0.359-0.450) [0.345-0.463]  & (0.357-0.443)  [0.343-0.457] \\
  $R_t$          &         0.927 $\pm$ 0.061      &   0.900 $\pm$ 0.050          \\
                 &   (0.806-1.048) [0.767-1.086]  & (0.802-0.998)  [0.771-1.029] \\
  $\Delta M_s$   &         17.3$^{+2.2}_{-1.3}$   &   18.0$^{+1.7}_{-1.5}$       \\
  ($ps^{-1}$)    &   (15.0-23.0)   [11.9-31.9]    & (15.4-21.7)    [14.8-25.9]   \\
$\vtd~(10^{-3})$ &          8.36 $\pm$ 0.55       &        8.15 $\pm$ 0.41       \\
                 &   (7.14-9.50)   [6.27-10.00]    & (7.34-8.97)    [7.08-9.22]   \\
  $\vtd/\vts$    &          0.209 $\pm$ 0.014     &      0.205 $\pm$ 0.011       \\
                 &   (0.179-0.238) [0.157-0.252]  & (0.184-0.227)  [0.177-0.233] \\
  $Im \lambda_t$ &         13.5 $\pm$ 1.2         &    13.04 $\pm$ 0.94          \\
    ($10^{-5}$)  &   (10.9-15.9)   [9.4-16.6]     & (11.2-14.9)    [10.6-15.5]   \\
\hline
\end{tabular}
\caption[]{ \small { Values and errors for different quantities from 
\cite{BUPAST}.
In brackets the 95$\%$ and 99$\%$ probability regions are also given.
$\lambda_t=V_{ts}^*V_{td}$.}}
\label{mfv}
\end{center}
\end{table}

\section{New Set of Fundamental Flavour Variables}
During the 1970's and 1980's the variables $\alpha_{QED}$, the Fermi 
constant $G_F$ and the sine of the Weinberg angle ($\sin\theta_W$) were 
the fundamental parameters in terms of which the electroweak tests of 
the SM have been performed. After the $Z^0$ boson has been discovered
and its mass precisely measured at LEP-I, $\sin\theta_W$ has been replaced
by $M_Z$ and the fundamental set used in the electroweak precision studies 
in the 1990's has been $(\alpha_{QED},G_F,M_Z)$. It is to be expected that
when $M_W$ will be measured precisely this set will be changed to 
$(\alpha_{QED},M_W,M_Z)$ or ($G_F,M_W,M_Z)$.

One can anticipate an analogous development  in this decade 
in connection with the CKM matrix. While the set (\ref{2.76}) has clearly 
many virtues and has been used extensively in the literature, one should
emphasize that presently no direct independent measurements of $\eta$
and $\varrho$ are available. $\eta$ can be measured cleanly in
the decay $K_L\to\pi^0\nu\bar\nu$. On the other hand to our knowledge
there does not exist any strategy for a clean independent measurement 
of $\varrho$. 

\begin{figure}[htb!]
\begin{center}
{\epsfig{figure=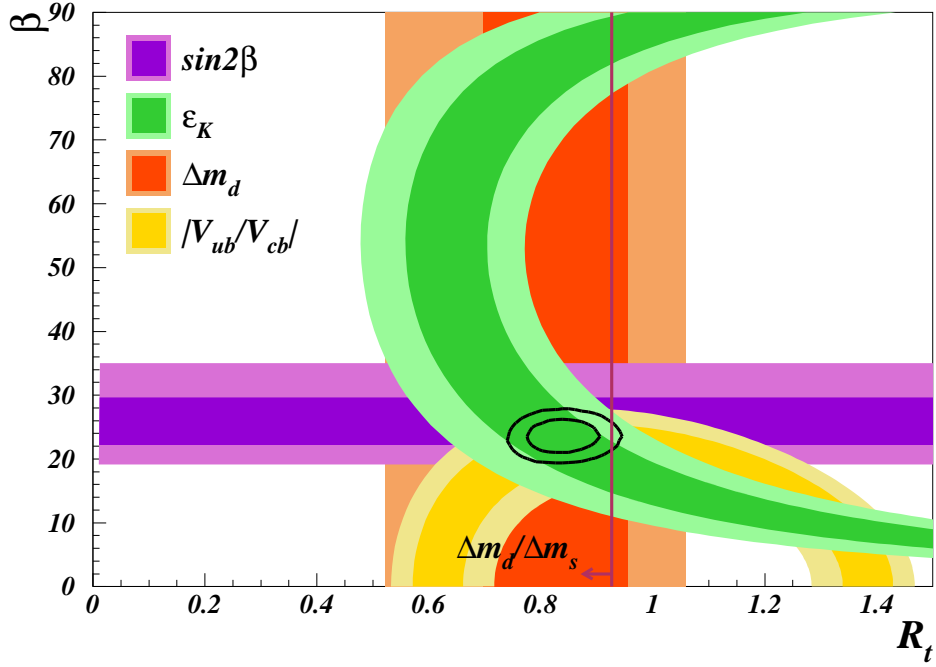,height=10cm}}
\caption[]{The allowed regions {(68$\%$ and 95$\%$)} 
in the ($R_t,\beta$) plane \cite{BUPAST}.
Different constraints are also shown.}
\label{fig:rtbeta}
\end{center}
\end{figure}

Taking into account the experimental feasibility of various measurements
and their theoretical cleanness, the most obvious candidate for the 
fundamental set 
in the quark flavour physics for the coming years 
appears to be \cite{BUPAST}
\begin{equation}\label{I2}
\vus, \qquad \vcb, \qquad R_t, \qquad \beta
\end{equation}
with the last two variables measured by means of (\ref{Rt}) and 
(\ref{asy}), respectively.
In this context one can investigate, in analogy to the 
$(\bar\varrho,\bar\eta)$ 
plane, the $(R_t,\beta)$
plane for the exhibition of various constraints on the CKM matrix. 
We show this in fig.~\ref{fig:rtbeta}. Moreover inserting 
\be\label{S1}
\bar\varrho=1-R_t\cos\beta,\qquad \bar\eta=R_t\sin\beta
\ee
into (\ref{f1})-(\ref{2.83d}) and using
(\ref{RbRt}) it is an easy matter to express all elements of the CKM 
matrix in terms of the variables in (\ref{I2}):
\be
V_{ud}=1-\frac{1}{2}\lambda^2-\frac{1}{8}\lambda^4 +\ord(\lambda^6),
\qquad
V_{ub}=\frac{\lambda}{1-\lambda^2/2}\vcb \left[1-R_t e^{i\beta}\right],
\ee
\be
V_{cd}=-\lambda+\frac{1}{2} \lambda \vcb^2 -
\lambda\vcb^2 \left[1-R_t e^{-i\beta}\right] +
\ord(\lambda^7),
\ee
\be
V_{cs}= 1-\frac{1}{2}\lambda^2-\frac{1}{8}\lambda^4 -\frac{1}{2} \vcb^2
 +\ord(\lambda^6),
\ee
\be
V_{tb}=1-\frac{1}{2} \vcb^2+\ord(\lambda^6),
\qquad
V_{td}=\lambda\vcb R_t e^{-i\beta}
+\ord (\lambda^7),
\ee
\begin{equation}
 V_{ts}= -\vcb +\frac{1}{2} \lambda^2 \vcb - 
\lambda^2 \vcb \left[1-R_t e^{-i\beta}\right]
  +\ord(\lambda^6)~,
\end{equation}
where in order to 
simplify the notation we have used $\lambda$ instead of $\vus$ as
$V_{us}=\lambda+\ord(\lambda^7)$.  

For the fundamental set of parameters in the quark flavour physics given in 
(\ref{I2}) we have presently within the SM \cite{BUPAST}
\begin{equation}\label{FUND}
\vus=0.221\pm 0.002, \qquad \vcb=(40.4\pm0.8)\cdot 10^{-3}, \qquad 
R_t=0.90\pm0.05, \qquad \beta=(23.2 \pm 1.4)^\circ
\end{equation}
where the errors represent one standard deviations and the small shift in 
$\vcb$ results from the UT fit. The first entry will be soon replaced by 
$0.2241\pm 0.0036$ \cite{VUS}.  

In the future the situation may change and other sets of fundamental 
flavour variables could turn out to be more useful than the set (\ref{I2}). 
As argued in \cite{BUPAST}, replacing $R_t$ in (\ref{I2}) by $\gamma$ could 
result in the most useful set of flavour variables provided $\gamma$ 
can be precisely measured. Similarly the pair $(R_b,\gamma)$ is very useful 
as it gives the length of the hand of the unitarity clock in fig. \ref{L:2} 
and its position. 
Other possibilities are discussed in \cite{BUPAST}.

\section{Outlook: Shopping List}
The coming nine years should be very exciting in the field of flavour 
and CP violation due to a vast amount of data expected from laboratories 
in Europe, USA and Japan. One should also hope that theorists will 
sharpen their tools. There are already many reviews of the methods for the
extraction of the sides and angles of the UT 
\cite{BABAR,LHCB,FERMILAB,BF97,Nir}. Therefore I will  
be very brief here.

{\bf 1.} It is very desirable that the uncertainties in all inputs 
entering the five steps of the standard analysis of UT are reduced. 
The elements $|V_{ub}|$ and $\vcb$ play here a special role as they 
are essentially independent of possble new physics contributions. The 
improved accuracy on $\xi$ in (\ref{Rt}) together with a precise measurement
of $\Delta M_s$ will give us an accurate value of $R_t$ and consequently
by means of (\ref{VUBG}) a prediction for $\gamma$. However, the importance
of accurate values for $\hat B_K$, $F_{B_d}\sqrt{\hat B_d}$, 
and $F_{B_s}\sqrt{\hat B_s}$ should not be underestimated. These three 
quantities are easier to calculate than hadronic matrix elements relevant 
for non-leptonic K and B decays and are equally important. The precise 
knowledge of $\hat B_K$ combined
with improved accuracy on $\vcb$ will allow to use the precise value of 
$\varepsilon_K$ (step 3) more efficiently. 
An improved value of $F_{B_d}\sqrt{\hat B_d}$ 
combined with a more accurate value of $\mt$ will give us as seen in
(\ref{VT}) a precise value of $\vtd$ and consequently $R_t$. A precise 
value of $F_{B_s}\sqrt{\hat B_s}$ is important for other reasons. As 
$\vts$ is fixed by the CKM unitarity to be very close to $\vcb$, the 
measurement of $\Delta M_s$ combined with $F_{B_s}\sqrt{\hat B_s}$ 
allows the measurement of the box diagram function $S_0(x_t)$:
\begin{equation}\label{So}
S_0(x_t)=
2.32\left[
\frac{270\mev}{\sqrt{\hat B_{B_s}}F_{B_s}}\right]^2
\left[\frac{0.040}{|V_{ts}|} \right]^2
\left[\frac{\Delta M_s}{18.0/{\rm ps}} \right] 
\left[\frac{0.55}{\eta_B}\right]
\end{equation}
and consequently of $\mt$ that could be compared with its direct
measurement. This could teach us about the possible new physics 
beyond SM. For $m_t=167\pm 5$ GeV one has $S_0(x_t)=2.39\pm0.12$. 

{\bf 2.} The measurement of $\sin 2\beta$ by means of $a_{\psi K_S}(t)$ 
will certainly be improved in the coming years so that the angle $\beta$
will be known with an error of $1^\circ$ ! At this accuracy a closer 
look at possible theoretical uncertainties will be required. This very 
precise value for $\beta$ will be one day confronted with its value 
determined by means of clean decays $K_L\to\pi^0\nu\bar\nu$ and
$K^+\to\pi^+\nu\bar\nu$ \cite{BBSIN}. With the accuracy for both branching 
ratios 
of $10\%$ a measurement of $\sin 2\beta$ with an error $\pm 0.05$ 
becomes possible. In order to do better not only the accuracy 
on the branching ratios has to improve but also an NNLO QCD-analysis
of these decays combined with improved value of $\overline{m}_c(m_c)$
is required.

In the meantime the CP asymmetry in $B_d\to \phi K_S$ that also 
measures $\sin 2\beta$ will be one of the important topics. Being 
dominated by QCD penguin diagrams it is expected to be more sensitive 
to new physics than $a_{\psi K_S}(t)$. The first results from BaBar 
and Belle indicate a value for $\sin 2\beta$ that differs significantly from
(\ref{ga}). The recent excitement  about this anomaly could be 
premature, 
however, as the experimental errors are still large and the decay is
not as theoretically clean as $B\to \psi K_S$. Recent summary is given in 
\cite{NIR02}. See also \cite{PSIKS}.

{\bf 3.} Another hot topic is the measurement of $\sin 2\alpha$
through the CP asymmetry in $B_d\to \pi^+\pi^-$ that unfortunately is 
polluted by
QCD penguin diagrams and consequently by hadronic uncertainties. 
There is a vast literature on this subject and 
many suggestions have been put forward in order to overcome the hadronic 
uncertainties with the hope to extract the true angle $\alpha$. 
Unfortunately the BaBar and Belle data on $a_{CP}(\pi^+\pi^-)$ disagree 
with each other with the asymmetry being consistent with zero and 
large, respectively. Similarly there is no real consensus among theorists.
Recent summary is given in 
\cite{NIR02}. See also \cite{ALPHA}.
The situation reminds us of $\epe$ at the beginning of the 1990ïs. Yet, 
I am convinced that here the experimentalists will reach much faster the 
agreement than was the case of $\epe$. Moreover, as the theoretical 
issues appear to be less involved than in $\epe$, I expect that some
consensus will be reached by theorists in the coming years. On the 
other hand I have some doubts that a precise value of $\alpha$ will 
follow in a foreseable future from this enterprise.
However, one should also stress \cite{BBSIN} that only a 
moderately precise measurement of $\sin 2\alpha$ can be as useful for 
the UT as a precise measurement of the angle $\beta$. This has been recently 
reemphasized in \cite{BBNS2,BUPAST}. This is clear from table~\ref{mfv} that
shows very large uncertainties in the indirect determination of 
$\sin 2\alpha$. 

{\bf 4.} In view of the comments made in the previous section a precise 
measurement of the angle $\gamma$ is of utmost importance. An excellent 
overview 
of various strategies for $\gamma$ can be found in \cite{FLgamma}. 
The present efforts concentrate around the decays $B^0_d\to \pi K$ 
and $B^\pm\to \pi K$. On the one hand the data from BaBar and Belle 
improved considerably this year. On the other hand, there exist several
methods like QCDF \cite{QCDF} and PQCD \cite{PQCD} approaches and more 
phenomenological 
approaches: the mixed strategy \cite{FM}, the charged strategy
\cite{NRBOUND},  the neutral strategy \cite{GPAR3}
and the Wick contraction method \cite{IWICK,BSWICK}. While I agree to some 
extent with the Rome
group \cite{CIFRMAPISI} that the issue is more involved than stated 
sometimes by some authors, 
one
cannot deny a great progress made by theorists during the last three 
years and I am confident that a combination of all $B\to \pi K$ and 
$B\to\pi\pi$ channels will offer in due time a useful, if not the
most precise, determination of $\gamma$.

Another, very interesting line of attack is to use the U-spin 
symmetry \cite{RF99,RF991,GRCW,G00} for the determination of $\gamma$. 
In particular the strategies involving
the U-spin related decays $B^0_{d,s}\to \psi K_S$ and 
$B^0_{d,s}\to D^+_{d,s} D^-_{d,s}$
 \cite{RF99} and 
 $B^0_s\to K^+ K^-$ and $B^0_d\to\pi^+\pi^-$
\cite{RF991} appear to be promising for
Run II at FNAL and in particular for LHC-B. They are unaffected by FSI and 
are only limited by U-spin breaking effects. 

Yet, there is no doubt that at the end the most precise determinations of 
$\gamma$ will come from the strategies involving $B_d\to D^{(*)\pm}\pi^\mp$ 
\cite{DPI} and $B_s\to D_s^{(*)\pm}K^{(*)\mp}$ \cite{adk} in which all 
hadronic uncertainties cancel.
One should also mention the triangle construction of Gronau and Wyler 
\cite{Wyler} that uses $B^\pm\to K^\pm\{D^0,\bar D^0,D^0_\pm\}$ 
where $D^0_\pm$ 
denotes the CP eigenstates of the neutral $D$ system.  However, this method 
is problematic because of the small
branching ratios of the colour supressed channel $B^{+}\to D^0 K^{+}$
and its charge conjugate.
Variants of this method
which could be more promising have been proposed in \cite{DUN2,V97}.

{\bf 5.} Finally a few rare K and  B decays should be put on this
shopping list. 
The recent events for $K^+\to\pi^+\nu\bar\nu$ are very encouraging 
\cite{KPDATA}.
In particular one can construct the UT exclusively by 
means of $K_L\to\pi^0\nu\bar\nu$ and $K^+\to\pi^+\nu\bar\nu$ \cite{BBSIN}. 
See also \cite{NirGro}.
The 
accuracy of this construction can compete with the one by means of B 
decays, provided the branching ratios are precisely measured and the
uncertainties in $\vcb$ and $m_c$ reduced. Similarly
$K_L\to\pi^0\nu\bar\nu$ appears to be the best decay to measure the 
area of the unrescaled unitarity triangle or equivalently $\imlt$ in 
table 1. Finally 
\begin{equation}\label{bxnn}
\frac{Br(B\to X_d\nu\bar\nu)}{Br(B\to X_s\nu\bar\nu)}=
\left|\frac{V_{td}}{V_{ts}}\right|^2, \qquad
\frac{Br(B_d\to\mu^+\mu^-)}{Br(B_s\to\mu^+\mu^-)}=
\frac{\tau_{B_d}}{\tau_{B_s}}\frac{m_{B_d}}{m_{B_s}}
\frac{F^2_{B_d}}{F^2_{B_s}}
\left|\frac{V_{td}}{V_{ts}}\right|^2
\end{equation}
allow to determine $\vtd/\vts$  or equivalently $R_t$ that can be 
compared with its determination by means of $\Delta M_d/\Delta M_s$ 
in (\ref{Rt}). 
As these decays are dominated in the SM by $Z^0$-penguin diagrams, 
while $\Delta M_{d,s}$ are governed by box diagrams, this comparision 
offers a very good test of the SM. 

\section{Going Beyond the Standard Model}
If the SM is the proper description of flavour and CP violation, all 
branching ratios and CP asymmetries are given just in terms of four 
flavour variables, such as the sets (\ref{2.76}), (\ref{I2}) or the sets 
considered in \cite{BUPAST}. This necessarily implies relations between
various branching ratios and asymmetries that have to be satisfied
independently of the values of the flavour parameters in question if
the SM is the whole story. Such relations have been extensively 
studied in \cite{BBSIN,BF01,REL,UUT}.

Now, beyond the SM the amplitude for any decay can be generally written as
\cite{Pisa}
\be\label{master}
{\rm A(Decay)}= \sum_i B_i \eta^i_{\rm QCD}V^i_{\rm CKM} 
\lbrack F^i_{\rm SM}+F^i_{\rm New}\rbrack +
\sum_k B_k^{\rm New} \lbrack\eta^k_{\rm QCD}\rbrack^{\rm New} V^k_{\rm New} 
\lbrack G^k_{\rm New}\rbrack\, .
\ee
The non-perturbative parameters $B_i$ represent the hadronic matrix elements 
of relevant local 
operators $Q_i$  present in the SM. For instance in the case of 
$K^0-\bar K^0$ mixing, the matrix element of the operator
$\bar s \gamma_\mu(1-\gamma_5) d \otimes \bar s \gamma^\mu(1-\gamma_5) d $
is represented by the parameter $\hat B_K$ in (\ref{100}).
There are other non-perturbative parameters in the SM that represent 
matrix elements of operators $Q_i$ with different colour and Dirac 
structures. 

The objects $\eta^i_{\rm QCD}$ are the QCD factors analogous to $\eta_i$ and 
$\eta_B$.
Finally, $F^i_{\rm SM}$ stand for 
the so-called Inami-Lim functions \cite{IL} that result from the calculations 
of various
box and penguin diagrams. They depend on the top-quark mass. An example is
the function $S_0$ in (\ref{100}).

New physics can contribute to our master formula in two ways. First, it can 
modify the importance of a given operator, that is relevant already in the SM, 
through the new short distance functions $F^i_{\rm New}$ that depend on 
the new 
parameters in the extensions of the SM like the masses of charginos, 
squarks, charged Higgs particles and $\tan\beta=v_2/v_1$ in the MSSM. 
These new 
particles enter the new box and penguin diagrams. Secondly, in more 
complicated extensions of the SM new operators (Dirac structures) that are 
either absent or very strongly suppressed in the SM, can become important. 
Their contributions are described by the second sum in 
(\ref{master}) with 
$B_k^{\rm New}, \lbrack\eta^k_{\rm QCD}\rbrack^{\rm New}, V^k_{\rm New}, 
G^k_{\rm New}$
being analogs of the corresponding objects in the first sum of the master 
formula. The $V^k_{\rm New}$ show explicitly that the second sum describes 
generally new sources of flavour and CP violation beyond the CKM matrix. 
This sum may, however, also include contributions governed by the CKM 
matrix that are strongly suppressed in the SM but become important in 
some extensions of the SM. A typical example is the enhancement of the 
operators with Dirac structures $(V-A)\otimes(V+A)$, 
$(S-P)\otimes (S\pm P)$ and 
$\sigma_{\mu\nu} (S-P) \otimes \sigma^{\mu\nu} (S-P)$ contributing to 
$K^0-\bar K^0$ and $B_{d,s}^0-\bar B_{d,s}^0$ mixings in the MSSM with large 
$\tan\beta$ and in supersymmetric extensions with new flavour violation. 
The latter may arise from the misalignement of quark and squark mass
matrices. 

Now, the new functions $F^i_{\rm New}$ and $G^k_{\rm New}$ as well as the 
factors $V^k_{\rm New}$ may depend on new CP violating phases complicating 
considerably phenomenological analysis. 
On the other hand there exists a class of extensions of the SM in which
the second sum in (\ref{master}) is absent 
(no new operators) and flavour changing transitions are 
governed by the CKM matrix. In particular there are no new complex 
phases beyond the CKM phase. We will call this scenario ``Minimal Flavour 
Violation" (MFV) \cite{UUT} being aware of the fact that for some authors 
MFV means a more general framework in which also new operators can give 
significant contributions. 
See for instance the recent discussions in 
\cite{BOEWKRUR,AMGIISST}.
In the MFV models, as defined here,  the  master formula (\ref{master}) 
simplifies to 
\be\label{mmaster}
{\rm A(Decay)}= \sum_i B_i \eta^i_{\rm QCD}V^i_{\rm CKM} 
\lbrack F^i_{\rm SM}+F^i_{\rm New}\rbrack 
\ee 
with $F^i_{\rm SM}$ and $F^i_{\rm New}$ being real. 

Many relations between various quantities valid in the SM are also valid for 
MFV models or can be straightforwardly generalized to these models.
One of the interesting properties of the MFV models is the existence of 
the universal unitarity triangle (UUT) \cite{UUT} that can be constructed 
from quantities in which all the dependence on new physics cancels out
or is negligible like in tree level decays from which $|V_{ub}|$ and 
$|V_{cb}|$ are extracted. The values of $\bar\varrho$, $\bar\eta$, $\alpha$,
$\beta$, $\gamma$, $R_b$, and $R_t$  resulting from 
this determination are
the ``true" values that are universal within the MFV models. 
Various strategies for the determination of the UUT are discussed 
in \cite{UUT}.

The presently available quantities that do not depend on the new physics 
parameters within the MFV-models and therefore can be used to determine 
the UUT are $R_t$ from $\Delta M_d/\Delta M_s$ by means of (\ref{Rt}),
$R_b$ from $\vub$ by means of (\ref{2.94})   and $\sin 2\beta$ 
extracted from the CP asymmetry in $B^0_d\to \psi K_S$. 
Using only these three quantities, we show in figure \ref{fig:figmfv}  
the allowed universal region for $(\bar\varrho,\bar\eta)$ (the larger
ellipse) in the MFV models 
\cite{BUPAST}. The central values, the errors and 
the $95\%$ (and $99\%$)  C.L. ranges for various quantities of 
interest related to this UUT are collected in table \ref{mfv}.
Similar analysis has been done in \cite{AMGIISST}.

It should be stressed that any MFV model that is inconsistent with the 
broader allowed region in figure \ref{fig:figmfv} and 
the UUT column in table \ref{mfv} is ruled out. 
We observe that there is little room for MFV models that in their predictions 
differ significantly from the SM. It is also clear that to 
distinguish the SM from the MFV models considered here on the 
basis of the analysis of the UT only, will require considerable reduction of 
theoretical uncertainties.

Such a distinction should be much easier in the MSSM with minimal flavour 
violation but with large $\tan\beta$. Even if the CKM parameters in this model
are expected to be very close to the ones in the SM, the presence of 
 neutral Higgs boson penguin-like diagrams with charginos
and stop-quarks in the loop can
{\it increase} by orders of magnitude the
branching ratios of the rare decays
$B^0_{s,d}\to\mu^+\mu^-$~\cite{BAKO,CHSL,HULIYAZH,BOEWKRUR} and to
{\it decrease} significantly the $B^0_s$-$\bar B^0_s$ mass difference
$\Delta M_s$~\cite{BUCHROSL1} relative to the expectations based on
the SM. Most recent discussions, including also more complicated scenarios, 
can be found in \cite{BOEWKRUR,BUCHROSL1,NEW,AMGIISST,DEPI}. 

\section{Concluding Remarks}
The recent direct measurements of $\sin 2\beta$ by BaBar and Belle opened 
a new era of the precise tests of the flavour structure of the SM and its 
extensions. These measurements have shown how important it is to have 
quantities that are free of theoretical uncertainties. With a single direct 
and clean measurement of the angle $\beta$ it was possible to achieve 
accuracy comparable with the indirect measurements of $\beta$ that involves 
simultaneously a number of quantities like $|V_{ub}|$, $\vcb$, $\Delta
M_{s,d}$ and $\varepsilon_K$ that are all subject to theoretical uncertainties.
This lesson makes it clear that one should make all efforts to realize the 
clean strategies that involve $B_d\to D^{(*)\pm}\pi^\mp$ 
and $B_s\to D_s^{(*)\pm}K^{(*)\mp}$ for $\gamma$, $K\to\pi\nu\bar\nu$ for 
$\sin 2\beta$, $Im\lambda_t$ and the UT as well as the rare decays 
$B\to X_{s,d} \nu\bar\nu$ and $B_{s,d}\to\mu^+\mu^-$ relevant for 
$\vtd/\vts$. Similar comments apply to a number of strategies with small 
uncertainies as $\Delta M_d/\Delta M_s$ discussed in section 7. However, to 
this end our theoretical tools have to be improved.

The next years will certainly bring new insight in the flavour structure of 
the SM and its extensions. In particular it will be important to resolve 
the issues related to CP asymmetries in $B_d\to\pi^+\pi^-$ and 
$B_d\to\phi K_S$ and to measure $\Delta M_s$. Similarly it will be important 
to see whether large $\tan\beta$ effects predicted in the MSSM in 
$B_{s,d}\to\mu^+\mu^-$ and $\Delta M_s$ are realized in nature. As emphasized
by Peccei \cite{Peccei} even more important is the search for new complex 
phases with 
the hope to find convincing scenarios that would simultaneously explain the 
the size of CP violation in the low enery processes and the matter-antimatter 
asymmetry observed in the universe. Another issue is the CP violation in the 
neutrino sector.

It is conceivable that the physics responsible for 
the matter-antimatter asymmetry involves only very short distance scales,
as the GUT scale or 
the Planck scale, and the related CP violation is unobservable in the
experiments performed by humans. Yet even if  such an unfortunate situation
is a real possibility, it is unlikely that the single phase in the CKM 
matrix provides a fully adequate description of CP violation at scales 
accessible to experiments
peformed on our planet in this millennium. On the one hand the
KM picture of CP violation  is so economical that it
is hard to believe that it will pass future experimental tests in spite of 
its recent successes seen in fig. 4. On
the other hand almost any extention of the SM contains
additional sources of CP violating effects \cite{CHRO}. 
As some kind of new physics
is required in order to understand the patterns of quark and lepton
masses and mixings and generally to understand the flavour dynamics, 
it is very likely that this physics will bring
new sources of CP violation modifying KM picture considerably.
In any case the flavour physics and CP violation will remain to be a
very important and exciting field at least for the next 10-15 years.

\section*{Acknowledgments}
I would like to thank Fabrizio Parodi and Achille Stocchi for a 
wonderful collaboration on the matters related to the UT and
the organizers for inviting me to such a pleasant 
conference. I also than the Max-Planck Institute for Physics in Munich for
covering my travel expenses. The work presented here
 has also been supported 
in part by the German Bundesministerium f\"ur Bildung und Forschung under 
the contract 05HT1WOA3 and the DFG Project Bu. 706/1-1.


\begin{thebibliography}{99}
\bibitem{WARN} 
The numerical value on the r.h.s of (\ref{100}) being proportional to
$\lambda^{-10}$ is very sensitive to $\lambda$ but as $A\propto \lambda^{-2}$
and $P_c(\varepsilon)\propto \lambda^{-4}$ the constraint (\ref{100}) depends
only quadratically on $\lambda$.
\end{thebibliography}
\end{document}